# Optical writing of Dzyaloshinskii Néel-type skyrmions

Dinar Khusyainov[1*], Lukas Körber[1], Sheng Li[1], Rein Liefferink[1], Thibaud Denneulin[2], András Kovács[2], Fabian Kammerbauer[3], Robert Frömter[3], Mathias Kläui[3], Dmytro Afanasiev[1], Alexey Kimel[1], Johan H. Mentink[1], Rafal E. Dunin-Borkowski[2], Theo Rasing[1]

[1*] *Radboud University, Institute for Molecules and Materials, Nijmegen, 6525 AJ, Netherlands.*
[2]*Ernst Ruska-Centre for Microscopy and Spectroscopy with Electrons, Forschungszentrum Jülich, Jülich, 52428, Germany.*
[3]*Institute of Physics, Johannes Gutenberg University Mainz, Mainz, 55099, Germany.*

*Corresponding author. E-mail:dinar.khusyainov@ru.nl;

Ultrashort laser pulses offer a route to write nanoscale topological magnetic textures for information processing. Magnetic skyrmions are prototypical examples, yet optically generating them with controlled position, type, and, in particular, chirality remains challenging. This challenge arises because magnetic dipolar interactions, which often drive skyrmion formation, preserve the degeneracy between opposite chiralities, leaving no uniquely selected spin configuration. Here, we show that in an ultrathin Co/Pt film, sequences of picosecond circularly polarized laser pulses can write Néel-type, fixed-chirality skyrmions stabilized solely by the Dzyaloshinskii–Moriya interaction. Combining Lorentz transmission electron microscopy and magneto-optical microscopy, we comprehensively reconstruct their internal spin structure across multiple length scales, ranging from nanometres to micrometers. Using asymmetric skyrmion-expansion measurements, we determined the sign and magnitude of the Dzyaloshinskii–Moriya interaction and identified the written chirality. Our findings connect deterministic optical writing with the controlled spin structure of skyrmions, advancing optical control of topological textures for opto-spintronic applications.

## Introduction

Ultrashort laser pulses enable writing magnetic textures for nanoscale data storage on ever shorter time scales [1–7]. Recent experiments have demonstrated laser-induced formation of a variety of magnetic structures, ranging from magnetic domain walls [8] to exotic topological states such as meron–antimeron pairs [9], bimerons [10], and even three-dimensional hopfions [11]. Among these textures, skyrmions are particularly appealing as potential information carriers for racetrack memories [12, 13] and reservoir computing[14, 15]. Inside these nanoscale magnetic textures, the magnetization twists continuously across space, forming a topologically distinct configuration characterized by an integer topological charge [16–18] and a well-defined chirality that determines the sense of magnetization rotation. One of the most important interactions defining skyrmion chirality is the Dzyaloshinskii–Moriya interaction (DMI), which becomes particularly important in thin ferromagnetic films where dipolar interactions are suppressed [19, 20]. There, DMI stabilizes Néel-type skyrmions with a preferred chirality set by its sign, which we refer to as Dzyaloshinskii Néel-type skyrmions[21].

Despite extensive demonstrations of laser-induced skyrmion formation in various magnetic systems, including magnetic multilayers [2, 5, 6, 22, 23, 4, 24, 25], the optical writing of Dzyaloshinskii Néel-type skyrmions remains unexplored. More generally, on-demand nucleation and control over the internal spin structure of laser-induced skyrmions remain challenging. Previous studies have shown that skyrmions generated by short laser pulses typically form at random positions within the excitation area [5, 25, 26], requiring additional techniques such as nanolithography [27] or ion irradiation [22, 23, 28] to localize their nucleation. In ultrathin ferromagnetic films with perpendicular magnetic anisotropy, helicity-dependent all-optical switching (HD-AOS) enables local magnetization reversal using sequences of circularly polarized laser pulses[29–33] and has been shown to generate nanoscale domains that may serve as skyrmion nuclei [33, 34]. This raises the possibility of using HD-AOS to control skyrmion nucleation. However, it is unclear whether this approach can reliably produce skyrmions with a well-defined internal spin structure and DMI-favoured chirality. Conventional magneto-optical imaging primarily probes the out-of-plane magnetization and therefore cannot unambiguously distinguish spin textures or determine their chirality. Lorentz transmission electron microscopy (LTEM) in Fresnel mode and off-axis electron holography provides a powerful alternative, enabling direct visualization of in-plane magnetization components in one-dimensional [35], two-dimensional [5, 36–38], and even three-dimensional solitonic structures [39, 40]. Therefore, applying LTEM to study helicity-dependent laser-induced domain formation in ultrathin ferromagnetic films offers a promising pathway toward resolving open questions about the spin structure of laser-induced magnetic skyrmions in ultrathin ferromagnetic films.

Here, we investigate the optical nucleation of Dzyaloshinskii skyrmions in ultrathin Co/Pt layers. Combining magneto-optical microscopy and LTEM, we show that sequences of circularly polarized picosecond laser pulses produce Néel-type skyrmions with well-defined chirality, spanning from a few hundred nanometers to tens of μm in diameter. Their nucleation can be stochastic or deterministic, depending on the background magnetization and laser fluence. Asymmetric skyrmion-expansion measurements establish the sign of the DMI and identify the preferred chirality. We further estimate the threshold DMI strength required to stabilize Dzyaloshinskii skyrmions in thin ferromagnetic films. Our findings connect laser-induced domain nucleation with the internal spin structure of chiral magnetic textures and establish a framework for optically writing and identifying skyrmions whose chirality is selected by DMI.

## Results and discussions

We use a ferromagnetic multilayered structure, consisting of thin films of Ta (1.0 nm)/MgO (2.0 nm)/[Pt (3.0 nm)/Co (0.85 nm)]$_3$/Pt (3.0 nm)/Ta (5.0 nm) deposited on a 50 nm thick SiN membrane. The sample exhibits perpendicular magnetic anisotropy, with an out-of-plane magnetic hysteresis loop and a coercive field of 24 mT (See supplementary material 1 and Methods). We first determine the parameters for helicity-dependent all-optical switching in the sample using magneto-optical Faraday microscopy (See Fig. 1a). In our experiments, we vary the fluence, the number of pulses, and the magnetization direction of the initial monodomain state, and capture images of the nucleated magnetic domains after illumination with circularly polarized 3-ps-long pulses laser pulses. For multipulse excitation, we use a train of pulses separated by 2 ms. Four magnetic states were observed: monodomain, deterministic, stochastic, and a mixed state of deterministic and stochastic switching (consistent with recent work[33]). Figure 1b shows background-subtracted magneto-optical images after illumination with a sequence of 200 laser pulses, as a function of fluence and magnetization orientation in the photo-irradiated area. The fluence threshold for nucleation of laser-induced domains, with a sequence of 200 pulses, is shifted by ~0.1 mJ/cm$^2$ for magnetization "up" (red) compared with magnetization "down" (blue). This result aligns well with observations of magnetic circular dichroism in thin ferromagnetic films[30–33, 41, 42]. It is demonstrated that a fluence of 1.83 mJ/cm² reliably initiates the formation of small, homogeneous domains in a deterministic manner from an initial state with the magnetization oriented "up," while we do not induce any domains starting from an initial magnetization "down". At a fluence of 2.00 mJ/cm² and initial magnetization "down", stochastic nucleation of small magnetic domains occurs within the illuminated area. In contrast, when the magnetization is flipped, the system exhibits a mixed state with stochastic domains at the center of the reversed spot and deterministic switching in the outer region. Supplementary Material 2 presents the complete diagram of magnetic states as a function of fluence and number of pulses.

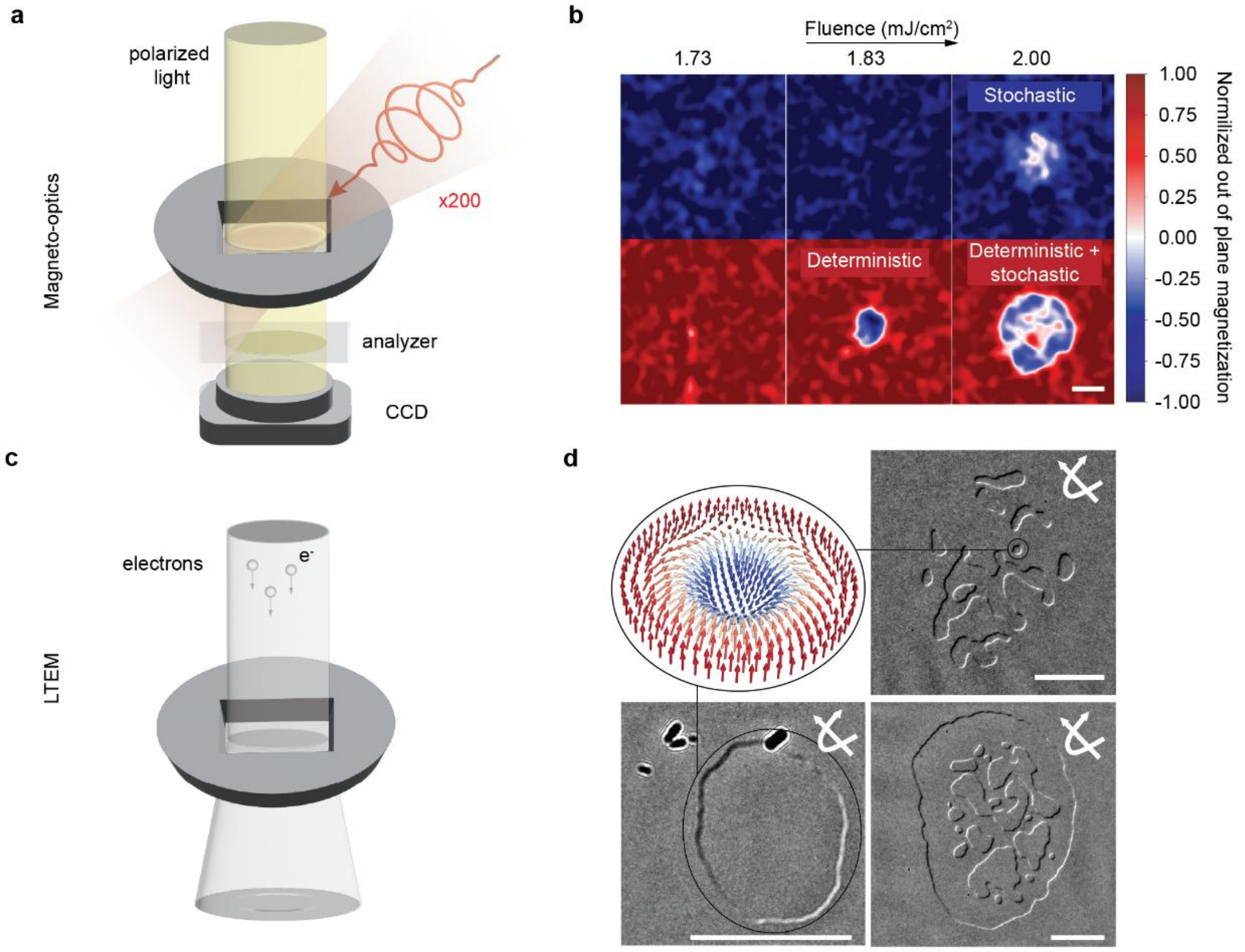

**Fig. 1. Optical writing of Dzyaloshinskii skyrmions.** (**a**) Sketch of the magneto-optical Faraday microscopy coupled with a source of picosecond circularly polarized laser pulses. The sample is grown on a 50-nm-thick SiN membrane window. (**b**) Magneto-optical images of the laser-induced domains. We plot magneto-optical images after illumination with 200 circularly polarized laser pulses at 3 fluences (1.76, 1.88, and 1.96 mJ/cm2) and two initial monodomain out-of-plane magnetization states, "up" (red) and "down" (blue). We background-subtracted and normalized the magneto-optical images and applied a Gaussian filter (σ=3) to reduce noise. (**c**) Sketch of a Lorentz transmission electron microscopy setup. (**d**) Fresnel defocus LTEM images of the laser-induced domains observed in magneto-optical images in (**b**). Images were taken at a tilt of 30º and a defocus of -15 mm. We processed the images with polynomial background subtraction to remove the background. Scale bars are 10 μm.

To improve spatial resolution and access in-plane magnetic induction contrast, we employed Lorentz transmission electron microscopy (LTEM). Domains induced by laser irradiation were produced under parameters consistent with those used in magneto-optical microscopy, and the samples were subsequently transferred to the

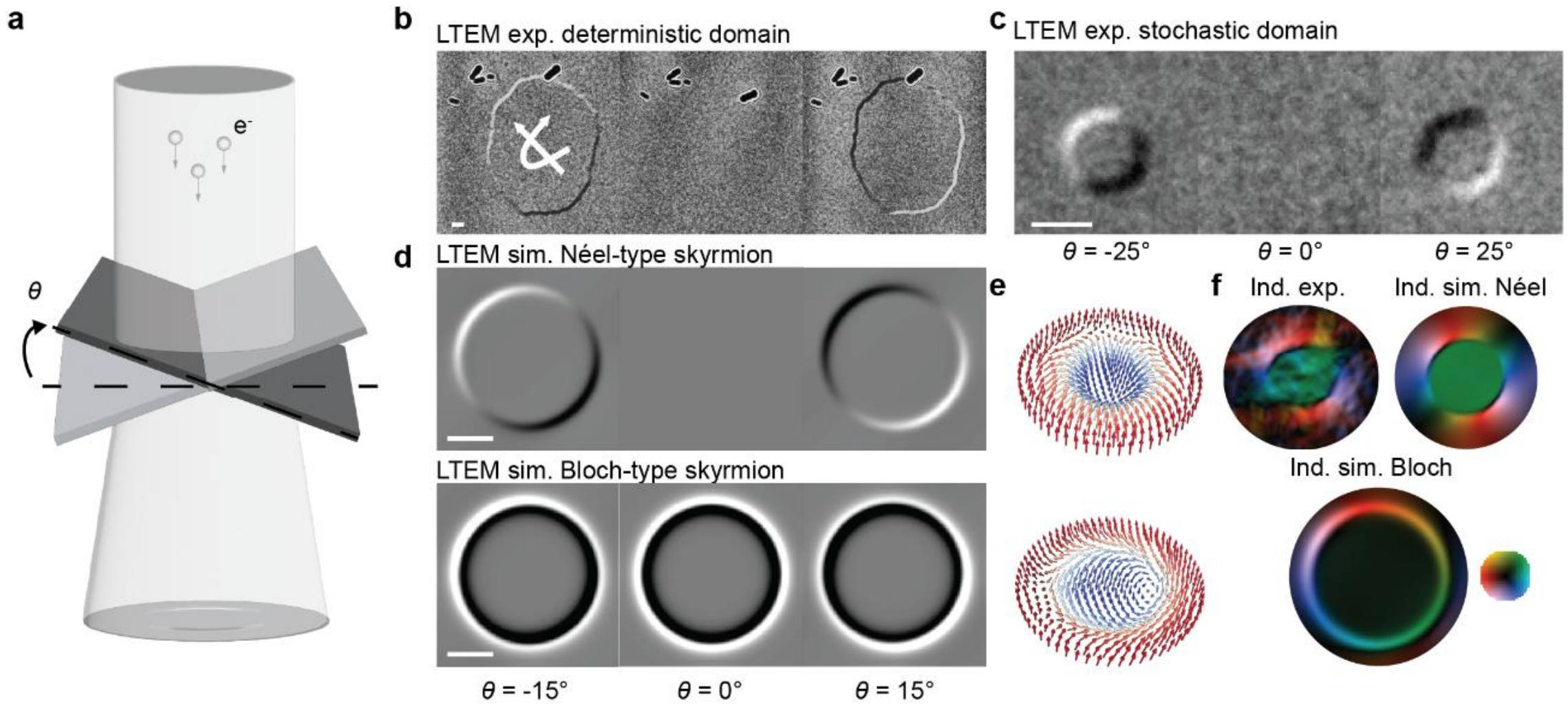

**Fig. 2. Identification of the type of magnetic skyrmion**. **(a)** Sketch of the experimental approach to measure tilt series using Fresnel defocus images in Lorentz mode. **(b)** Fresnel images of the tilt series of the deterministically laser-induced domain recorded at 3 tilt angles: -15, 0, 15º.We processed the images with polynomial background subtraction to remove the background. The scale bar is 1 μm. (**c**) Fresnel images of the tilt series of the stochastically laser-induced domain recorded at 3 tilt angles:- 25 °, 0 °, and 25 °. We processed the images with polynomial background subtraction to remove the background. The scale bar is 1 μm. (**d**) Simulated Fresnel LTEM images of the tilt series for Néel-type and Bloch-type skyrmions. The scale bar is 100 nm. (**e**) Sketch of spin configuration in Néel-type and Bloch-type skyrmions. (**f**) Reconstructed induction maps of the experimentally measured stochastic skyrmions using SITIE and the simulated Néel-type and Bloch-type skyrmions. The color wheel represents the in-plane direction of the magnetic induction.

LTEM system (see Fig. 1c). Within LTEM observations, magnetic domains corresponding to deterministic, stochastic, and mixed states are presented in Fig. 1d. The detection of magnetic domains required tilting the sample and setting high defocus values (-15 mm). Contrast was observed primarily at the domain walls (DWs), as LTEM signals arise from the interference of electron beams traveling through regions of in-plane magnetic domains with opposite magnetization. Although the magnetization within the domains is out of plane, tilting the sample causes contrast to emerge due to interference from the in-plane component of magnetization [35]. The contrast at the circular magnetic domain boundaries varies continuously from bright to dark. A line drawn through the regions of maximum bright and dark contrast corresponds to the tilt axis. The contrast pattern observed is indicative of Néel-type DWs (Fig. 1d). Laser-induced magnetic domains ranging from approximately 200 nm to 20 μm were observed. To verify the spin structure within the DWs, tilt-series measurements were performed on both deterministic µm-sized and stochastic nano-sized domains. Figure 2a illustrates the experimental setup used to acquire the tilt series. First, we recorded a tilt series of a deterministically created domain (Figure 2b). Three electron microscopy images were captured at tilt angles of -15°, 0°, and 15°. At 0° tilt, there is no contrast, whereas at tilt angles from -15° to 15°, the signal changes from bright to dark in the top-left part of the circular magnetic domain and from dark to bright in the bottom-right part. The tilt series was also conducted for the stochastically nucleated domains at tilt angles of -25°, 0°, and 25° (see Figure 2c). The observations indicate similar signals for the stochastically nucleated domains, suggesting that all studied laser-induced domains are the same.

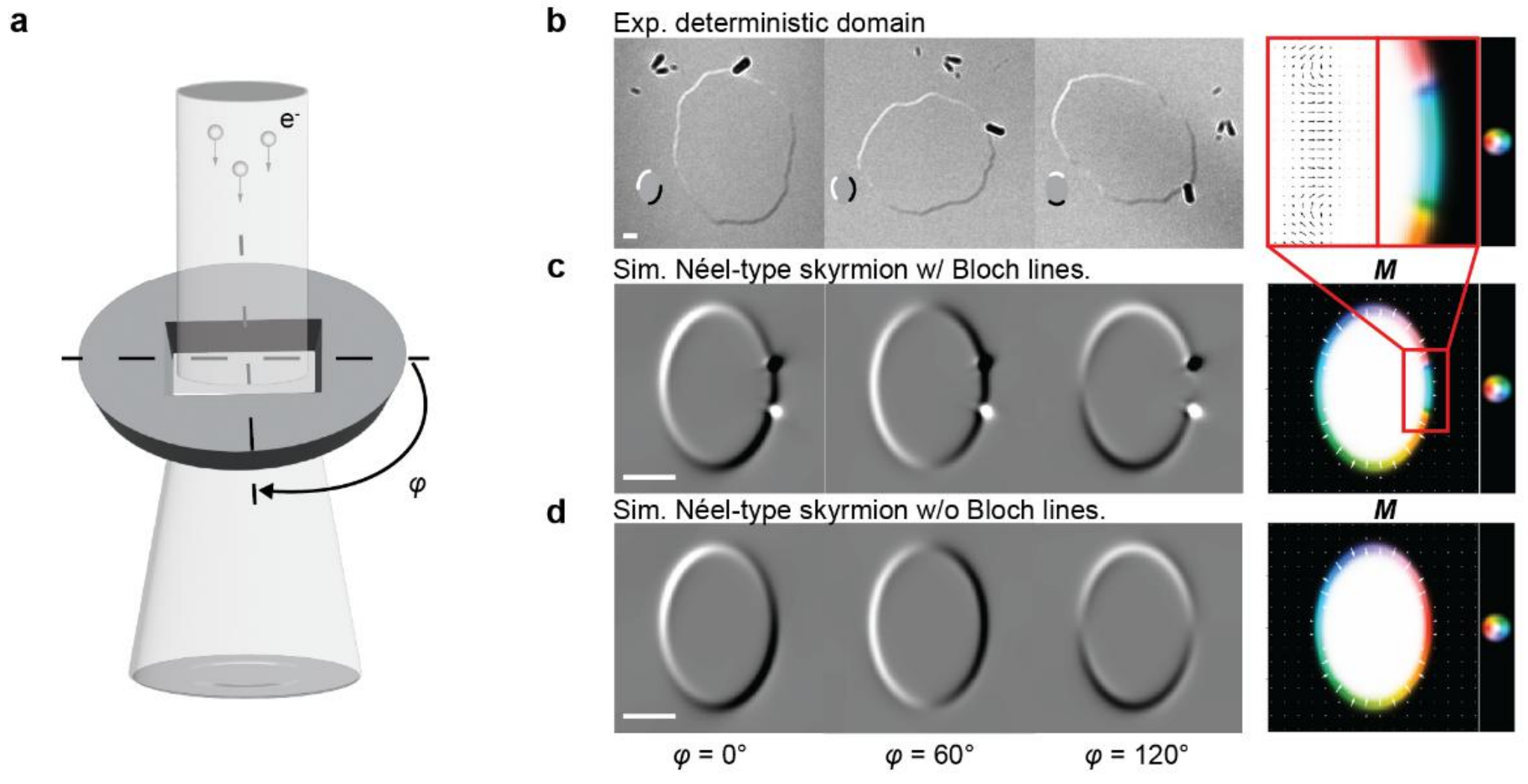

**Fig. 3 Azimuthal scan series of the deterministic laser-induced domain. (a)** Sketch of the experimental approach of recording LTEM images as a function of azimutal angle $\varphi$. (**b**) Azimuthal scan series of the LTEM images of the deterministic laser-induced domain as a function of azimuthal angle $\varphi$. The scale bar is 1 µm. **(c)** Simulated azimuthal series of LTEM images of the Néel-type skyrmion with defects in the domain wall for 3 azimuthal angles. We place two typical defects in the domain walls: Bloch lines. The zoom-in shows the spin configuration at the defects position. The color wheel represents the in-plane magnetization contrast. The scale bar is 100 nm (**d**). Simulated LTEM images: azimutal series of the Néel-type skyrmion without defects in the domain wall. The color wheel represents in-plane magnetization. The scale bar is 100 nm.

To prove the spin structure of the observed domains, we ran simulations of LTEM contrast. The simulations were performed for both Néel- and Bloch-type magnetic skyrmions, as depicted in Figure 2d. The LTEM images were simulated using a three-dimensional magnetization vector field corresponding to both Bloch-type and Néel-type skyrmions. These configurations correspond to idealized Bloch and Néel skyrmions (See Figure 2e). To compute the LTEM contrast, we use the PyLorentz software package (See Methods) [43]. The simulated LTEM contrast reveals notable differences between Bloch- and Néel-type skyrmions: (1) At zero tilt, Bloch-type skyrmions exhibit contrast, whereas Néel-type skyrmions do not; (2) At various tilt angles, Bloch-type skyrmions maintain consistent contrast, while Néel-type skyrmions demonstrate a marked reversal of contrast, transitioning from bright to dark at the top left of the skyrmion and from dark to bright at the bottom right of the skyrmion, corresponding to a reversal of the in-plane magnetization component due to the tilt; (3) The stronger Fresnel LTEM contrast of Bloch-type skyrmions arises from their in-plane magnetization configuration, which produces a stronger projected in-plane magnetic induction from that of Néel-type skyrmions[44].

The deterministically nucleated domain, owing to its size and the constraints imposed by the tilt limitation, exhibits nearly zero contrast or no apparent change in the region perpendicular to the tilt axis. Due to this uncertainty, in contrast, it is not possible to draw definitive conclusions regarding the topology of this domain at the indicated scale. Conversely, for small stochastic domains, no abrupt variations in contrast are observed along the DWs, enabling the plotting of induction maps using the transfer-of-intensity equation. A single-image transfer intensity equation approach is employed to generate the induction map of the small, stochastically nucleated domain[45]. The induction map for the experimentally observed stochastic domain is shown in Fig. 2f and compared with that of a simulated Néel-type skyrmion. This map further contrasts with the induction map produced by a simulated Bloch-type skyrmion. The simulated induction map of the Néel-type skyrmion clearly matches the experimentally optically written, randomly nucleated domain, indicating that Bloch-type contrasts are either absent or too faint to detect.

Subsequently, we examine the topological properties of the deterministically written domain. To confirm that it shares the same topological characteristics as the Néel-type skyrmion, it is essential to verify that the DWs along the domain are defect-free. This is achieved by rotating the sample about its azimuthal axis to accurately map the DWs of the deterministically written magnetic domain. Figure 3a illustrates the experimental approach involving azimuthal rotation of the sample. The sample is rotated azimuthally at a constant tilt angle. The results are depicted in Figure 3b for three distinct azimuthal angles. LTEM images reveal no defects in the DWs. To support this observation, we simulate an elliptical Néel-type skyrmion as in Fig. 2b and plot the simulated LTEM images as a function of azimutal angle $\varphi$. We perform the same simulations but account for two Bloch lines within the DW to imitate the most typical defects in a Néel-type DW. We present simulated LTEM images for the elliptical domain with Bloch lines in Figure 3c, demonstrating that defects in the DWs produce a significant signal and that a series of measurements at varying azimuthal angles can be used to map these defects. A comparison between simulation outcomes with and without defects in the DWs and experimentally measured results demonstrates that we do not observe any defects in the laser-induced deterministic domain. We conclude that a laser-induced deterministic magnetic domain has the same topology as a Néel-type skyrmion. However, LTEM does not allow us to access the chirality of the observed Néel-type skyrmions, as we see only the B-field distribution, which doesn't represent the in-plane configuration of spins within the DW. To determine the in-plane spin configuration within the DWs of laser-induced skyrmions and subsequently the sign of DMI stabilizing the chirality, we turn to magneto-optical microscopy.

To measure the sign of DMI and the corresponding chirality of the spin winding, we study asymmetric skyrmion expansion in an in-plane magnetic field. The direction of the expansion provides the DMI sign, indicating the chirality of the DW in the observed laser-induced magnetic skyrmions [46–48]. A sketch in Fig. 4a shows the experimental setup. We studied the sample using magneto-optical Kerr microscopy. To simplify the experiment, we nucleated magnetic skyrmions optically. The initial stage involved studying the symmetric skyrmion expansion caused by out-of-plane magnetic field pulses. An electromagnet pole was placed beneath the sample, and magnetic field pulses of ~40 mT, each lasting 0.1 s, were applied. We first nucleated a skyrmion (see Fig. 4b, top panel) with 200 pulses and expanded it using an out-of-plane magnetic pulse (See Fig. 4b, bottom panel). The symmetric radial skyrmion expansion was observed, as expected for a magnetic domain in an out-of-plane magnetic field. Next, we switched on an in-plane magnetic field of ~50 mT and applied an out-of-plane field pulse. The asymmetric expansion of the domain is shown in Fig. 4 c (bottom panel). The expansion is opposite to the direction of the in-plane magnetic field. To explain the observed asymmetric expansion, we turn to micromagnetic simulations (See Methods). In the micromagnetic simulations, the two distinct chiralities of the magnetic skyrmions were stabilized, corresponding to positive and negative DMI values. (See Fig. 4d). Then, an in-plane magnetic field of approximately 30 mT was applied to monitor the variations in the in-plane magnetization component of the simulated magnetic skyrmions. Fig. 4d displays the results. For a skyrmion with positive DMI (the in-plane magnetization component points toward the center), the preferred expansion direction under the applied in-plane magnetic field is opposite to the field. The Néel-type magnetic skyrmion with negative DMI shows the opposite behavior (the in-plane magnetization component extends outward from the center). In Supplementary Material 3, we estimated the DMI value using the asymmetric skyrmion expansion method, yielding $B_{\mathrm{DMI}} = 20 \pm 10$ mT. Our observed asymmetric skyrmion expansion is consistent with the case of positive DMI. This allows us to fully reconstruct the chiral spin structure of the laser-induced Néel-type skyrmion in our sample.

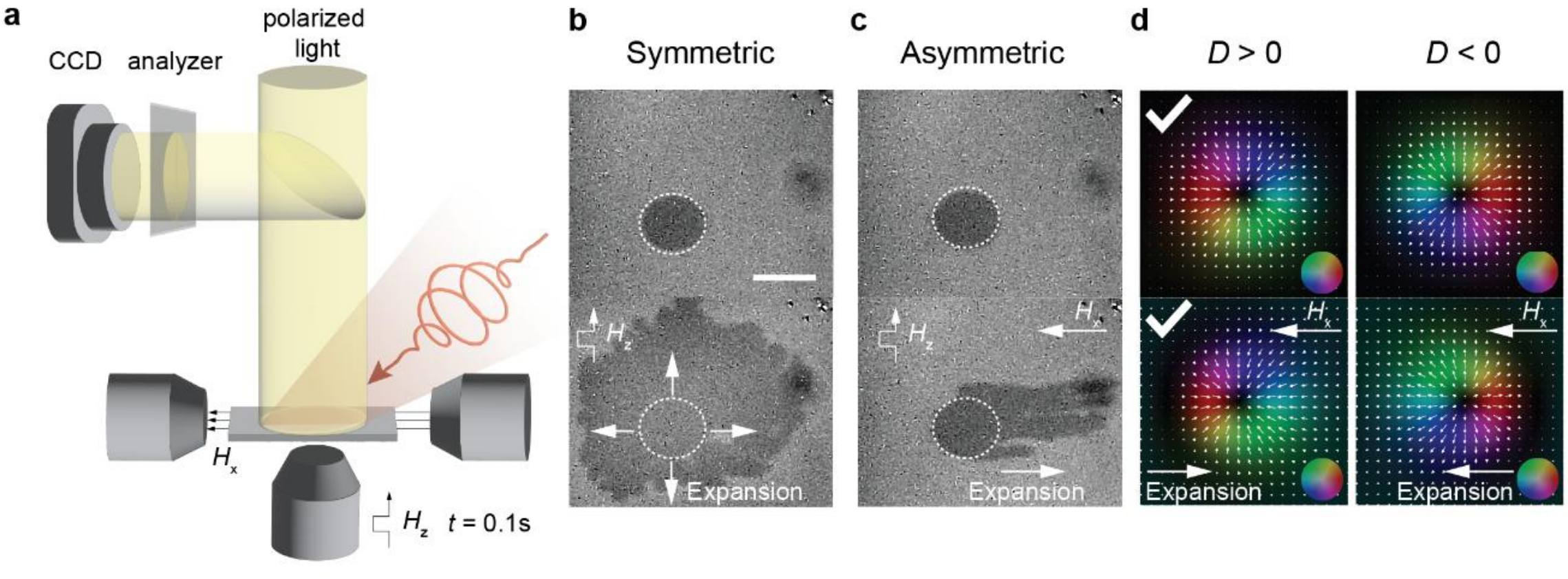

**Fig. 4. Determination of laser-induced Dzyaloshinskii skyrmion chirality. (a)** Sketch of the experimental setup for studying asymmetric skyrmion expansion. We use Magneto-optical Kerr microscopy and couple it with a source of short laser pulses. We use an electromagnet with a single pole to apply out-of-plane magnetic field pulses. We installed the sample between two electromagnet cores to provide an in-plane magnetic field. (**b**) Symmetric skyrmion expansion under the pulsed out-of-plane magnetic field. First, we nucleate a magnetic skyrmion using 200 laser pulses. In the bottom figure, we show the image after applying an out-of-plane magnetic field pulse of ~50 mT (c). Magneto-optical images of the laser-induced skyrmion (top). The bottom image shows asymmetric skyrmion expansion following an out-of-plane magnetic field pulse (33 mT, 0.1 s) while an in-plane magnetic field of ~50 mT is applied. We background-subtract the images using the line correction method. The scale bar is 30 µm. (**d**) Micromagnetic simulations of the Néel-type skyrmions with positive and negative DMI values demonstrating Néel-type skyrmions with opposite chirality. The bottom images demonstrate the in-plane magnetization component for both chiral skyrmions in the presence of an in-plane magnetic field of 30mT. The color wheel represents in-plane magnetization. We mark the case we observed in the experiment with a checkmark.

Our LTEM observations confirm that the domains in our ultrathin $[Co/Pt]_3$ system are Néel-type Dzyaloshinskii skyrmions. One can analytically estimate the minimum interfacial DMI required to completely suppress the Bloch counterparts of skyrmions by balancing the DMI against the dipole-dipole interactions. Following Thiaville et al. and Lemesh et al. [19, 49], this threshold parameter can be expressed as:

$$D_{\mathrm{thr}} \approx \frac{2\mu_0 M_s^2}{\frac{\pi^2}{t \ln 2}+\frac{\pi}{\Delta}}. \tag{1}$$

where $\mu_0$ is the vacuum magnetic permeability, $M_s$ is the saturation magnetization, $t$ is the thickness of magnetic layers, and $\Delta$ is a DW thickness. Applying an effective medium approach that accounts for the discrete layout of our thin magnetic layers and thick (3 nm) Platinum spacers, we estimate this critical threshold to be approximately $D_{\mathrm{thr}} = 0.2$ mJ/m$^2 < \Delta \cdot M_s \cdot B_{\mathrm{dmi}} = 0.22 \pm 0.11$ mJ/m$^2$ extracted from our skyrmion expansion measurements (and using parameters for $M_s = 1310$ kA/m and $\Delta = 8.6 \cdot$nm from [50]). The observed chirality of the skyrmions is consistent with DMI strength exceeding this threshold, within uncertainty, separating Dzyaloshinskii and hybrid skyrmions.

The Néel character of the DWs and the threshold DMI strength highlight the competition between DMI and dipolar interactions in determining the stability of magnetic textures. According to Eq. (1), $D_{\mathrm{thr}}$ increases almost linearly with the ferromagnetic layer thickness, $t$. Consequently, in multilayers with tens of Co/Pt repeats, the threshold quickly becomes unattainable[51]. In that case, the Bloch component of a DW can adopt either of two opposite chiralities, stabilizing hybrid skyrmions with uncontrollable chirality[5, 29]. Consequently, sufficiently long DWs are entropically guaranteed to exhibit boundaries between chirality, vertical Bloch lines, diminishing the wall robustness, and, in the case of skyrmions, neutralizing their topological charge. By contrast, when $D$ exceeds $D_{\mathrm{thr}}$, this degeneracy is strictly broken. Violating the preferred chirality instead costs a DMI energy penalty per unit length that counters any entropic gain. This is consistent with the observed defect-free homochirality.

As shown in Fig. 4 (further shown in Supplementary Material 3), asymmetry in skyrmion expansion does not change within the large area of the sample (hundreds of microns), meaning that DMI does not change locally, allowing us to conclude that chirality of laser-induced skyrmions will be controllable within the sample and stabilized by DMI. This control over chirality of optically written skyrmions is essential for the next generation of skyrmion devices, which could be provided using ultrathin ferromagnetic films hosting Dzyaloshinskii skyrmions.

# Conclusion

This work presents a framework for controllable writing of Néel-type Dzyaloshinskii skyrmions and for reconstructing the internal spin structure of laser-induced magnetic domains using LTEM and magneto-optical microscopy. Using this framework, we demonstrate controllable nucleation of Dzyaloshinskii skyrmions using sequences of circularly polarized laser pulses utilizing helicity-dependent all-optical switching. We can tune nucleation from deterministic to stochastic via external parameters such as background magnetization and fluence of circularly polarized laser pulses. We show that magnetic domains, whether nucleated deterministically or stochastically in an ultrathin ferromagnetic film with perpendicular magnetic anisotropy, share the same internal spin structure as Néel-type skyrmions. Experiments on asymmetric expansion allowed us to extract the positive DMI value and determine the chirality of optically written skyrmions. We estimated the threshold DMI value to be $0.22 \pm 0.11$ mJ/m$^2$, which is sufficient to stabilize the Dzyaloshinskii DWs and skyrmions in our ultrathin sample. Controlling the writing of skyrmion types and their spin structures creates opportunities for advances in skyrmion technology. Moreover, optically written Dzyaloshinskii skyrmions can accelerate the operational speed of skyrmion-based racetrack memory and reservoir computing devices.

# Methods

**Sample.** Ta (1.0 nm)/MgO (2.0 nm)/[Pt (3.0 nm)/Co (0.85 nm)]$_3$/Pt (3.0 nm)/Ta (5.0 nm) deposited on a 50 nm thick $Si_3N_4$ membrane. The silicon nitride membranes were grown on a 3-mm-diameter Si disk. 9 windows were etched in the Si disk (https://www.plano-em.de/). Magnetic properties were characterized using a Faraday rotation setup, with the out-of-plane hysteresis loop measured. The structure shows a coercive field of about 24 mT.

**Faraday/Kerr microscopy**

We used the magneto-optical Faraday effect to visualize the laser-induced domains in the Pt/Co/Pt trilayer. A white light source was used for illumination. We use thin-film polarizers for polarizer and analyzer. To magnify the image, we utilized a ×10 objective lens (10X Mitutoyo Plan Apo Infinity Corrected Long WD Objective). After the analyzer, the light was projected onto a charge-coupled device camera. The same microscope was reconfigured for Kerr geometry to collect light from the region near the membrane (See Fig. 4 a). The laser excitation beam was aligned in a non-collinear geometry and focused on the sample surface with a 250-mm-focal-length lens.

**Laser excitation.** We used an amplified Ti:sapphire laser system (Spectra-Physics Spitfire Ace) with a central wavelength of 800 nm, 3-picosecond pulses, and a repetition rate of 1 kHz to locally manipulate the magnetization of the sample surface. A quarter-wave plate was utilized to adjust the helicity of the laser pulses, while an external Pockels cell was employed to control the number of pulses during sequential illumination.

**LTEM.** The LTEM images were acquired using an aberration-corrected microscope (FEI Titan 80-300) operated at 300 kV. The magnetic-field-free environment (Lorentz mode) at the samples was achieved by turning the conventional objective lens off. The TEM images were recorded using a 2k x 2k resolution CCD camera (Gatan Ultrascan).

**Simulation of LTEM images**

To simulate LTEM images, we used the PyLorentz package (https://github.com/PyLorentz/PyLorentz/). As input, we prepared vector field plots for Bloch-type and Néel-type skyrmions. Simulations were performed for an electron beam energy of 300 keV. The electron beam was tilted along both the x- and y-directions to introduce sensitivity to in-plane magnetic components. Sample parameters were defined assuming a uniform thin film with a saturation magnetization of 1T and a thickness of 3 nm. The electrostatic potential was set to zero, neglecting thickness variations and electrostatic phase contributions. No supporting membrane was included in the model. Fresnel-mode LTEM images were simulated by propagating the electron wavefunction through the microscope using a contrast transfer formalism. Microscope parameters included a spherical aberration coefficient of 200 mm, an aperture semi-angle of $1.0 \times 10^{-5}$rad, and a defocus spread of 80 nm. Images were generated at a defocus value of $-100$ µm, corresponding to typical experimental Fresnel imaging conditions. No amorphous background scattering was included.

**Micromagnetic calculations**

Micromagnetic simulations were performed with the UBERMAG package using the OOMMF backend (oommfc 0.65.0, discretizedfield 0.91.0, micromagneticmodel 0.64.0) [52]. The magnetic stack was modeled as a dilute square monolayer with a lateral size of 200 nm and 8.55 nm thickness on a rectangular grid of 80x80x1 cells. The saturation magnetization was set to $\widetilde{M}_s = fM_s$ with a filling factor of $f \approx 0.3$ and $M_S = 1310$ kA/m, exchange stiffness $A_{ex} = 14\ \mathrm{pJ/m^3}$, interfacial DMI constant $D = \pm 1.2\ \mathrm{mJ/m^2}$ [50]**.** The anisotropy constant was fixed at the saturation field of $B_{sat} = 15$ mT. A static out-of-plane external field $B_z = 100$ mT was applied to stabilize the Skyrmion and mimic the effect of pinning. The initial Skyrmion was obtained by minimizing the total energy containing exchange, anisotropy, dipolar, Zeeman, and Dzyaloshinskii-Moriya interactions, starting from a nonchiral skyrmion. For the deformed Skyrmions, an additional in-plane field of an additional in-plane field of $B_z = \pm 30$ mT was applied.

# Acknowledgments

The authors thank Felix Büttner for fruitful discussion. The authors thank K. Saeedi and C. Berkhout for their technical support. This work was funded by the European Union Horizon 2020 and innovation program under the European Research Council ERC Grant Agreement No. 856538 (3D-MAGiC), ERC grant 485 101078206 ASTRAL, and the Gravitation program of the Dutch Ministry of Education, Culture and Science (OCW) under the research program “Materials for the Quantum Age” (QuMat) registration number 024.005.006. The group in Mainz acknowledges financial support from the Deutsche Forschungsgemeinschaft (DFG, German Research Foundation) SFB TRR 173 Spin+X (project A01, B02, A12 and B12 #268565370, TRR 288 Elasto-Q-Mat (project A12) #422213477, and the Horizon Europe project no. 101070290 (NIMFEIA) as well as TopDyn.

## Author contributions

D.Kh., J.M, and T. R. conceived the project. A. K., D.A., A.V. K., J.M, R. D., and T. R. supervised the study. D. Kh. and S.Li designed and built a Faraday and MOKE microscopy setup. D. Kh. and S.Li performed the magneto-optical experiments. D. Kh., A. K, T.D, S.Li measured LTEM images F. K., R. F. fabricated the samples under M.K.'s supervision. D. Kh. analyzed and interpreted the results of magneto-optical and LTEM experiments. D. Kh., T.D, L.K., R.L. carried out simulations. D. Kh., D.A., J.M., and T. R. wrote the original draft with feedback from all co-authors.

## Competing interests

The authors declare no competing interests.

# Supplementary Materials: Optical writing of Dzyaloshinskii Néel-type skyrmions.

Dinar Khusyainov[1*], Lucas Körber[1], Sheng Li[1], Rein Liefferink[1], Thibaud Denneulin[2], András Kovács[2], Fabian Kammerbauer[3], Robert Frömter[3], Mathias Kläui[3], Dmytro Afanasiev[1], Alexey V. Kimel[1] Johan H. Mentink[1], Rafal E. Dunin-Borkowski[2], Theo Rasing[1]

[1*] *Radboud University, Institute for Molecules and Materials, Nijmegen, 6525 AJ, Netherlands.*
[2]*Ernst Ruska-Centre for Microscopy and Spectroscopy with Electrons, Forschungszentrum Jülich, Jülich, 52428, Germany.*
[3]*Institute of Physics, Johannes Gutenberg University Mainz, Mainz, 55099, Germany.*

*Corresponding author. E-mail:dinar.khusyainov@ru.nl ;

# Table of Contents

# Supplementary Material I. Hysteresis loop.

We employed the Faraday microscopy to measure the sample's hysteresis loop shown in Fig S1. A HeNe laser served as the light source. We used a Glan-Taylor polarizer and analyzer for polarization control. The laser light was modulated with a photoelastic modulator and focused onto the sample's surface, which was placed between the poles of an electromagnet with openings in the cores. The magnetic field was oriented out-of-plane. The laser was transmitted through the poles and the sample, then was collimated by another lens. The beam subsequently passed through an analyzer and was detected by a single diode connected to a lock-in amplifier. The lock-in reference signal was derived from the photoelastic modulator controller.

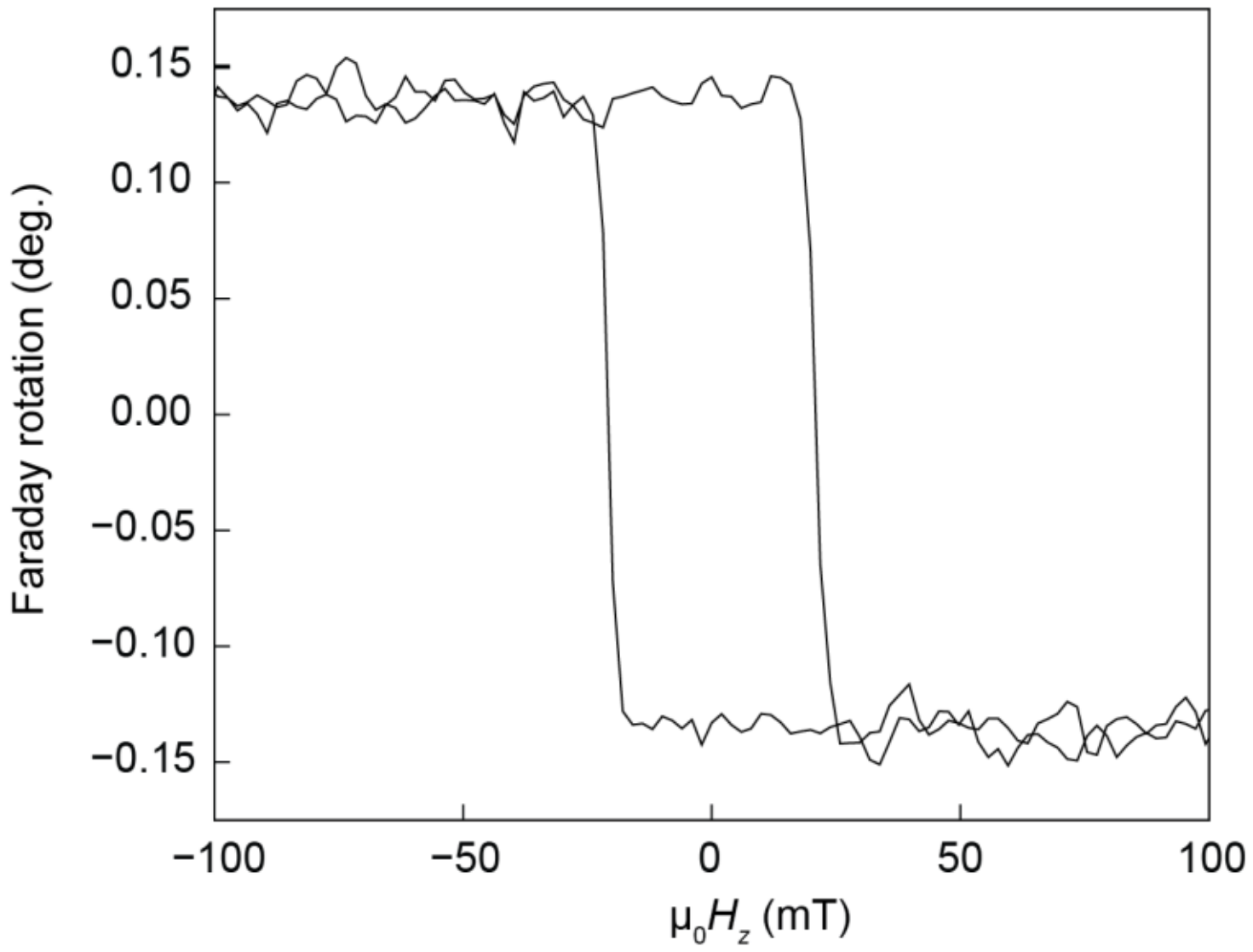

**Fig. S1**. **Hysteresis loop**. The hysteresis loop of Ta (1.0 nm)/MgO (2.0 nm)/[Pt (3.0 nm)/Co (0.85 nm)]$_3$/Pt (3.0 nm)/Ta (5.0 nm) was measured by observing the Faraday rotation in response to an external out-of-plane magnetic field.

## Supplementary Material II. Number of pulses – fluence phase diagram.

This supplementary information provides additional data for Figure 1b. It includes normalized magneto-optical Faraday microscopy images of laser-induced domains using 3 ps long circularly polarized laser pulses, shown as a function of number of pulses and fluence (See Fig. S2). Each image has been background-subtracted and processed with a Gaussian filter (sigma = 3). The results indicate that magnetization cannot be switched homogeneously and deterministically with a single pulse. Additionally, the nucleation threshold shifts to lower fluence levels, likely due to heat accumulation across pulses. Homogeneous and deterministic reversal of magnetic skyrmion bubbles was achieved at a fluence of 1.83 mJ/cm$^2$, starting from 50 to 200 pulses. When fluence exceeds this value, a deterministic ring forms, with stochastic nucleation occurring in the center. Additionally, we provide a comparison of the effects of left-circularly polarized pulses with diferent initial magnetization (“up” and “down”) to demonstrate that this effect is helicity-dependent (See Fig. S3).

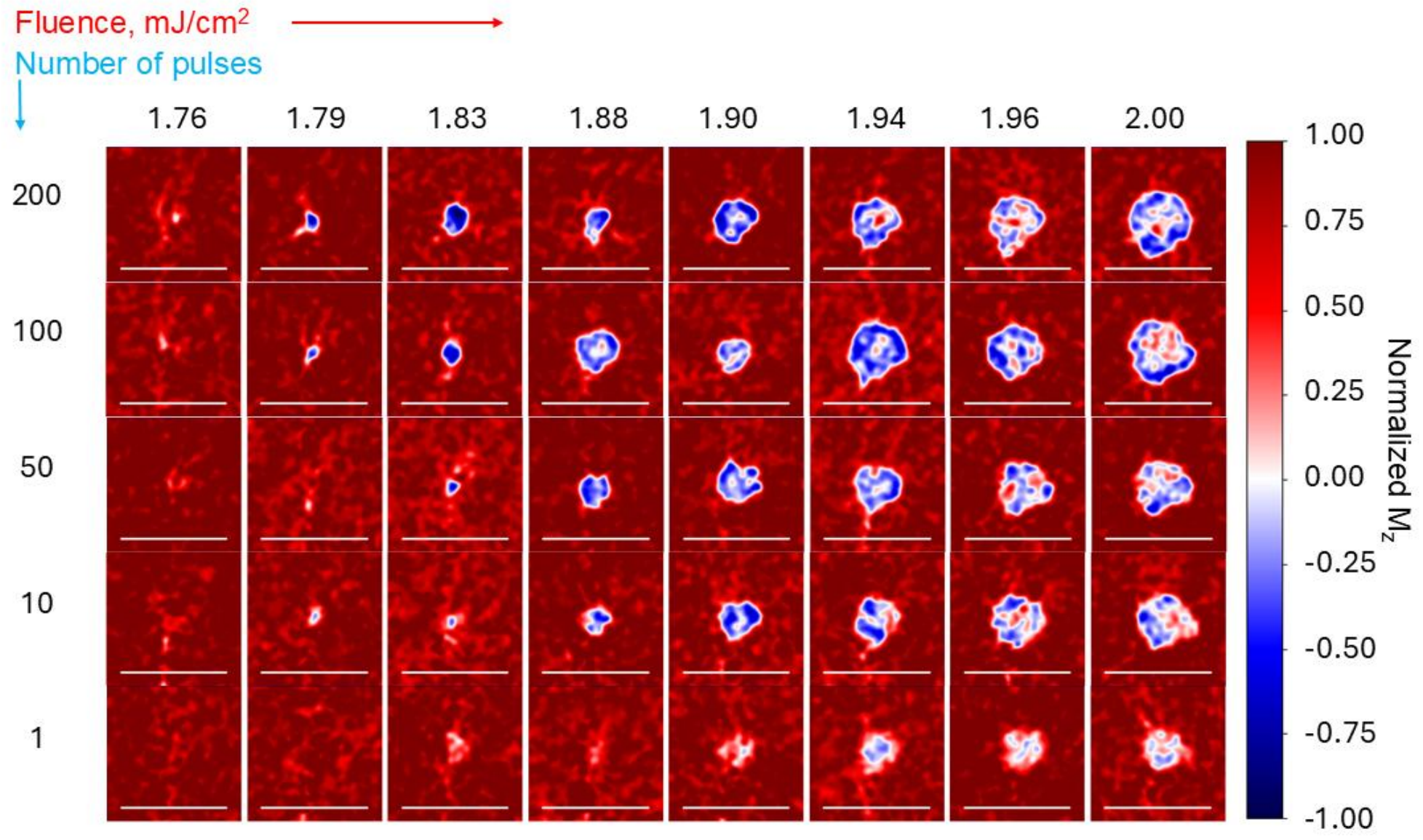

**Fig. S2**. **Deterministic switching conditions**. Normalized magneto-optical images of the laser-induced domains in the studied sample as a function of the number of circularly polarized laser pulses. All images are background-subtracted. We apply a Gaussian filter (sigma = 3). The scale bar is 50 µm.

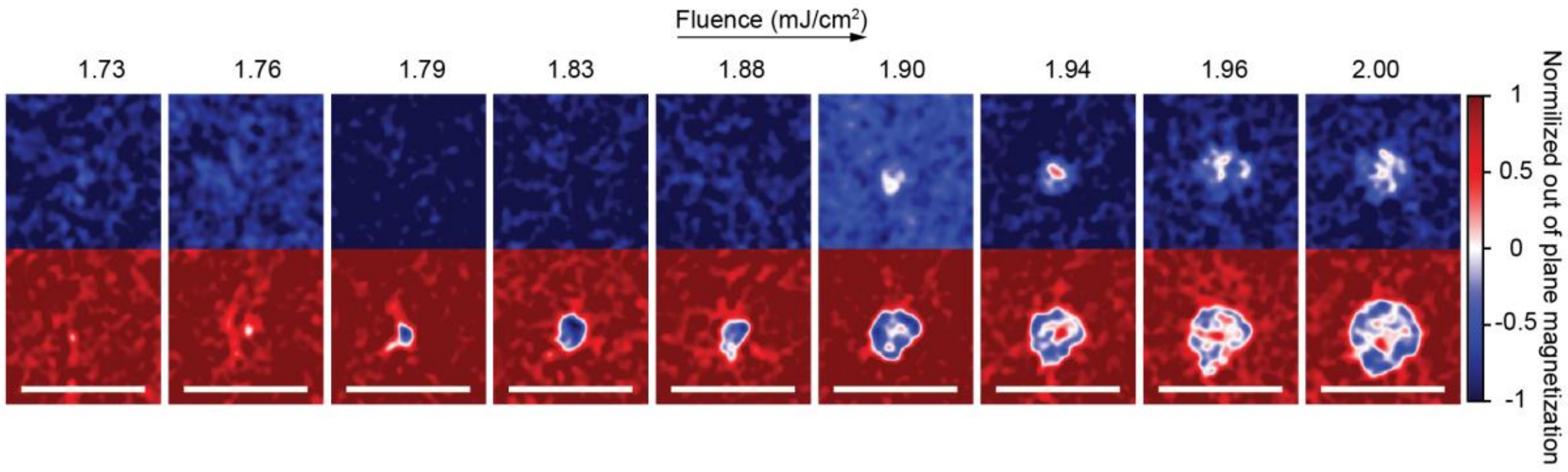

**Fig. S3**. **Helicity-dependent all-optical switching**. Background-subtracted magneto-optical images of the laser-induced domains with 200 pulses as a function of fluence and initial magnetization. We use left-circularly polarized light for both magnetizations. It can be seen that magnetization "up" couples with left-circularly polarized light, while magnetization "down" does not couple to light with this helicity. The scale bar is 50 µm.

## Supplementary Material III. Determination of the DMI value.

In this supplementary material, we estimate the DMI value by studying asymmetric skyrmion expansion in Kerr microscopy. We present the experimental results in Fig. S3. First, we nucleate a magnetic bubble deterministically. Next, we apply out-of-plane magnetic pulses of 33 mT for 0.1 s while varying the applied in-plane field. Corresponding images after applying each out-of-plane pulse are shown in Fig. S3 on the left side. We extract the speed of the asymmetric expansion to the right and left sides of the skyrmion and plot it in the graph. It can be seen that the left-side expansion reaches a minimum at around 20±10 mT. This minimum corresponds to the DMI value in the sample[1].

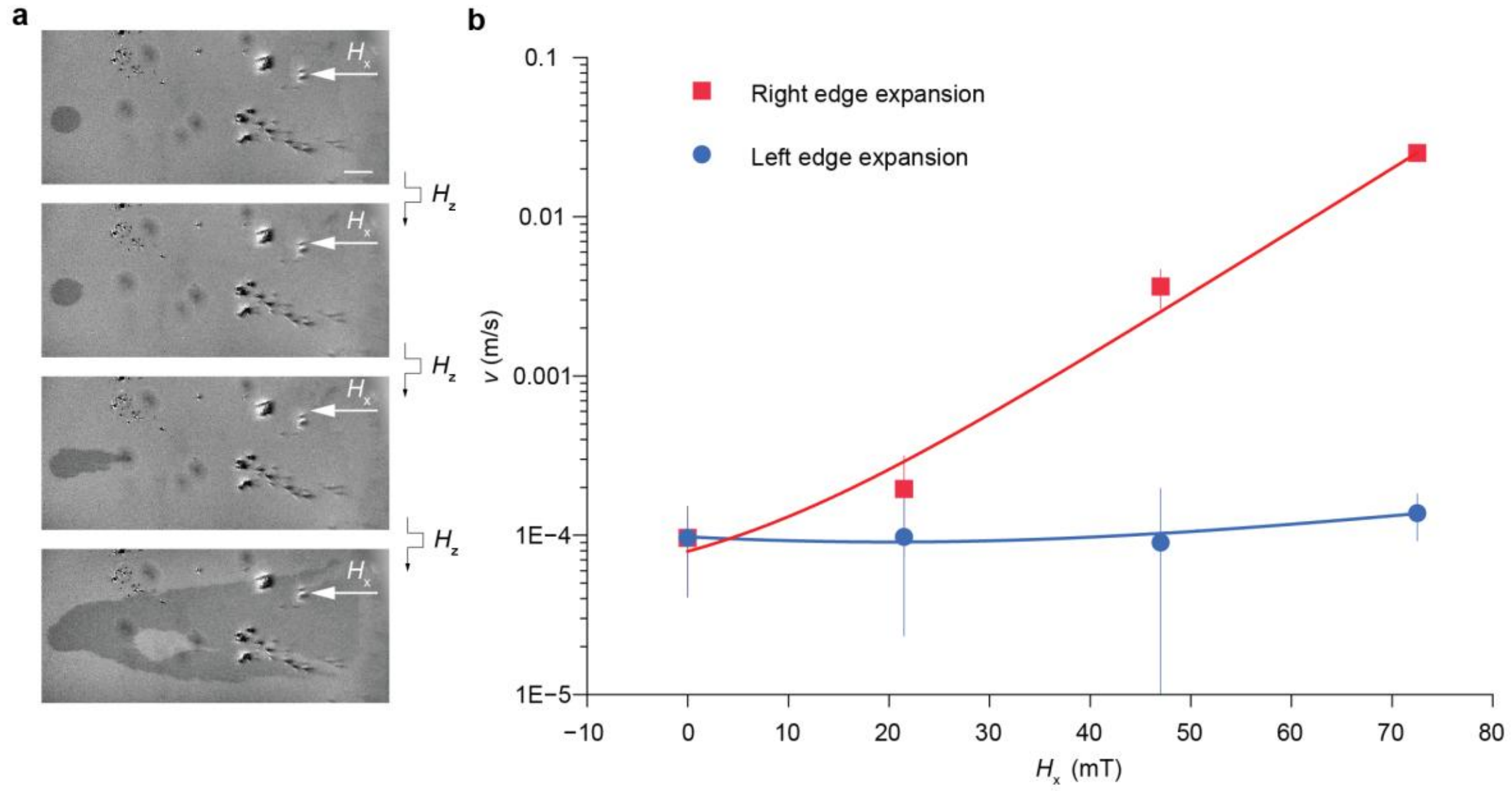

**Fig. S3**. **Determining the DMI value through asymmetric skyrmion expansion. (a)** We plot Magneto-optical Kerr microscopy images of the asymmetric bubble expansion after applying an out-of-plane magnetic pulse of $H_z$=33 mT for 0.1 s. We vary the in-plane magnetic field for each image. We background-subtract the images using the line-correction method. The scale bar is 30 µm. **(b)** Domain wall speed of the right and left edges of the magnetic bubble as a function of the in-plane magnetic field.